# Algorithmic Complexity in Real Financial Markets

R. Mansilla [1,2]

[1] **Center for Interdisciplinary Research in Science and The Humanities, National University of Mexico.**

[2] **Faculty of Mathematics and Computer Science, University of Havana, Cuba.**

**Abstract.**

A new approach to the understanding of complex behavior of financial markets index using tools from thermodynamics and statistical physics is developed. Physical complexity, a magnitude rooted in Kolmogorov-Chaitin theory is applied to binary sequences built up from real time series of financial markets indexes. The study is based on NASDAQ and Mexican IPC data. Different behaviors of this magnitude are shown when applied to the intervals of series placed before crashes and to intervals when no financial turbulence is observed. The connection between our results and The Efficient Market Hypothesis is discussed.

> "A theory is more impressive
>
> the greater the simplicity of
>
> its premises, the more different
>
> the kinds of things it relates and
>
> the more extended its range of applicability"
>
> **ALBERT EINSTEIN, 1949**

**1 Introduction.**

In the last years financial markets have received a growing attention from general public. Weather storms are discussed with the same emphasis in journal, newspapers and TV news than the financial ones and their endurance and consequences are analyzed by specialist in both field.

Physics have started few years ago to investigate financial data since they are remarkably well-defined complex system, continuously monitored down to time scale of seconds. Besides, almost every economic transaction is recorded, and an increasing amount of economic data is becoming accessible to the interested researchers. Hence, financial markets are extremely attractive for researcher aiming to understand complex systems as Mantegna and Stanley [1] stated in their book.

Several toy models of the markets behavior have been developed, highlighting among them the so-called Minority Game (MG) [2]. In that model, a magnitude resembling the real volatility (and also called *volatility*), have been extensively studied [2-10]. As we have proved in [11] for MG and in [12-13] for more sophisticated situations, there are more sensitive measures of the behavior of these models, which have their origin in statistical physics and thermodynamics. That is the so-called *physical complexity* [14-16] a magnitude rooted in the Kolmogorov-Chaitin theory [17-18]. In [12] we proposed an *ansatz* of the type $C(l) \approx l^a$ for the average value of the physical complexity taken over an ensemble of binary sequences. We also proved that the exponent $a$ strongly depend on the parameters of the model and can be seen as a measure of the coordination of agents in these models.

The aim of this work is to extend the study of this magnitude and his properties to the time series of real financial markets. To do that, we first develop a codification procedure, which translate real financial series in binary digits series. In spite of its *ad hoc* character and simplicity this procedure capture several important features of real financial markets series. As we will also prove, the behavior of the above mentioned measure of complexity drastically varies when applied to those intervals of the financial time series placed before the crashes and those where no financial turbulence is observed. We claim that this fact is close related with the Efficient Market Hypothesis (EMH) [19].

The structure of this paper is as follows: In Sec. 2 we discuss the way in which a real time series is encoded in a binary string. We also briefly develop the theoretical tools used in our analysis. In Sec. 3 we expose our results concerning the behavior of the physical complexity and in Sec. 4 we discuss our results in the framework of the most accepted paradigm of financial markets behavior.

**2 The binary string associated to the real time series.**

The MG is an eloquent even though rough description of financial markets. One could conceive the last institutions as a set of $N$ agents some of them taking a binary decision (buy=0, sell=1, for example) every time step. The main difference in this aspect with MG is that each agent take a decision in MG every time step meanwhile in real financial markets the agents often do not participate and prefer wait for a more clear position of the market [20].

The situation is schematized in Fig. 1. Let $N_0^t, N_1^t$ be the number of agents taking the decision $0,1$ respectively at time $t$. In MG we have $N_0^t + N_1^t = N$ for every time $t$, but in real financial market, as we remarked above, it is false. Hence in MG we only could have situations such as C or D, meanwhile in real financial market we have A or B in general.

It will be useful in our future discussion to establish the link between minority choice (no matter in what of the above scenarios) and the definition of price. To do it, let suppose that in the time $t_0$, $N_0^{t_0}$ agents decide to buy and $N_1^{t_0}$ decide to sell. Then if $N_0^{t_0} < N_1^{t_0}$ the winning choice corresponds to buy. Because there are less agents willing to buy than those willing to sell (supply exceeds demand) the price is pushed down. Notice that the smaller the ratio $N_0^{t_0}/N_1^{t_0}$ (recall it means lower price) the more overwhelming the victory of those in the minority room. On the other hand, if $N_0^{t_0} > N_1^{t_0}$ the winning choice correspond to sell. Now there are less agents willing to sell than those willing to buy (demand exceed supply) and the price is pushed up. Therefore, the larger the ratio $N_0^{t_0}/N_1^{t_0}$ the more overwhelming the victory of those in minority room again. The symmetric choice (buy=1, sell=0) also yields the same result if we consider the ratio $N_1^{t_0}/N_0^{t_0}$. Therefore the assumption that $p_t = p(N_0^t/N_1^t)$, that is, price is a monotonous increasing function of $N_0^{t_0}/N_1^{t_0}$ have been well established in the above discussion. The last statement is in perfect agreement with the situation shown in Fig. 1. All the points $(N_1, N_0)$ belonging to the same ray starting from the origin of coordinates represent different situations, but have the same ratio $N_0/N_1$, hence we may use a projective procedure and define $p_t = \arctan\left(\dfrac{N_0^t}{N_1^t}\right)$. We will not restrict our analysis to this ansatz, assuming for the function $p_t = p(N_0^t/N_1^t)$ a wider behavior.

The above discussions are also in agreement with some recent results [21] where crashes are studied as critical point. The key assumption of the above mentioned paper is that a crash may be caused by local self-reinforcing imitation between traders. If the agents tend to imitate each other, all may place the same order causing the crash. Bubbles can also be interpreted in this framework, as show us the "tulip affair" well described by B. G. Malkiel in [22].

We exploit all the above discussion to construct a binary series associated to each real time series of financial markets. The procedure is as follows: Let $\{p_t\}_{t \in N}$ be the original real time series. Then we construct a binary time series $\{a_t\}_{t \in N}$ by the rule:

$$a_t = \begin{cases} 1 & if \quad p_t \geq p_{t-1} \\ 0 & if \quad p_t < p_{t-1} \end{cases}$$

A similar technique was originally introduced in the context of natural languages [23]. More recently have been used in the study of cross-correlation between different market places [24] and in the study of the self-similarity in US dollar-German mark exchange rates [25]. To the outgoing binary series obtained by this procedure we apply the theoretical tools, which we briefly describe below. For a more complete discussion of these topics we suggest the readers [11], [12] and [16].

Physical complexity (first studied in [14] and [15]) is defined as the number of binary digits that are *explainable* (or meaningful) with respect to the environment in a string **h**. In reference to our problem the only physical record one gets is the binary string built up from the original time series and we consider it as environment $e$. We study the physical complexity of substrings of $e$. The comprehension of their complex features has high practical importance. The amount of data agents take into account in order to elaborate their choice is finite and of short range [22]. For every time step $t$ the binary digits $a_{t-l}, a_{t-l+1}, \ldots, a_{t-1}$ represent in some sense the winning choices made by the agents in the last $l$ time steps. Therefore, the binary strings $a_{t-l}, a_{t-l+1}, \ldots, a_{t-1}$ carry some information about the behavior of the agents. Hence, the complexity of these finite strings is a measure of how complex information agents face. We study the complexity of statistical ensembles of these substrings for several values of $l$.

We briefly review some measure devoted to analyze the complexity of binary strings. The Kolmogorov-Chaitin complexity [17], [18] is defined as the length of the shortest program $p$ producing the sequence $h$ when run on the universal Turing machine $T$:

$$K(h) = \min \{|p| : h = T(p)\} \qquad (1)$$

where $|p|$ represents the length of $p$ in bits, $T(p)$ the result of running $p$ on Turing machine $T$ and $K(h)$ the Kolmogorov-Chaitin complexity of the sequence $p$. In the framework of this theory, a string is said *regular* if $K(h) < |h|$. It means that $h$ can be described by a program $p$ with length smaller than the length of $h$.

As we have said, the interpretation of a string should be done in the framework of an environment. Hence, let imagine a Turing machine that takes the string $e$ as input. We can define the conditional complexity $K(h/e)$ [14-16] as the length of the smallest program that compute $h$ in a Turing machine having $e$ as input:

$$K(h/e) = \min \{|p| : h = C_T(p, e)\} \qquad (2)$$

We want to stress that $K(\mathbf{h}/\mathbf{e})$ represents those bits in $\mathbf{h}$ that are random with respect to $\mathbf{e}$ [14]. Finally, the physical complexity can be defined as the number of bits that are meaningful in $\mathbf{h}$ with respect to $\mathbf{e}$:

$$K(\mathbf{h}:\mathbf{e}) = |\mathbf{h}| - K(\mathbf{h}/\mathbf{e}) \qquad (3)$$

Notice that $|\mathbf{h}|$ represents (see [13-15]) the unconditional complexity of string $\mathbf{h}$ i.e., the value of complexity if the input would be $\mathbf{e} = \mathbf{f}$. Of course, the measure $K(\mathbf{h}:\mathbf{e})$ as defined in Eq. 3 has few practical applications, mainly because it is impossible to know the way in which information about $\mathbf{e}$ is encoded in $\mathbf{h}$. However (as shown in [16] and reference therein), if a statistical ensemble of strings is available to us, then the determination of complexity becomes an exercise in information theory. It can be proved that the average value $C(|\mathbf{h}|)$ of the physical complexity $K(\mathbf{h}:\mathbf{e})$ taken over an ensemble $\Sigma$ of strings of length $|\mathbf{h}|$ can be approximated by:

$$C(|\mathbf{h}|) = \langle K(\mathbf{h}:\mathbf{e}) \rangle_\Sigma \cong |\mathbf{h}| - K(\Sigma/\mathbf{e}) \qquad (4)$$

where:

$$K(\Sigma/\mathbf{e}) = -\sum_{\mathbf{h} \in \Sigma} p(\mathbf{h}/\mathbf{e}) \log_2 p(\mathbf{h}/\mathbf{e}) \qquad (5)$$

and the sum is taken over all strings $\mathbf{h}$ in the ensemble $\Sigma$. In a population of $N$ strings in environment $\mathbf{e}$, the quantity $\dfrac{n(\mathbf{h})}{N}$, where $n(s)$ denotes the number of strings equal to $\mathbf{h}$ in $\Sigma$, approximates $p(\mathbf{h}/\mathbf{e})$ as $N \to \infty$.

Let $\mathbf{e} = \{a_t\}_{t \in N}$ and $l$ a positive integer, $l \geq 2$. Let $\Sigma_l$ the ensemble of sequences of length $l$ built up by a moving window of length $l$ i.e., if $\mathbf{h} \in \Sigma_l$ then $\mathbf{h} = a_i a_{i+1} \cdots a_{i+l-1}$ for some value of $i$. We calculate the values of $C(l)$ using this kind of ensemble $\Sigma_l$. The selection of strings $e$ that we do in Sec. 3 is related to periods before crashes and in contrast, period with low uncertainty in the market.

**3 The study of real time series.**

We have used intraday series of the NASDAQ composite in the time period from January 3 1995 to April 18 2000 and also intraday series of IPC, the leader index of Mexican Stock Exchange (BMV) in the time period from January 2 1991 to March 27 2000. The evolution of both indices can be seen in Fig. 2. The numbers which appear in the horizontal axis of Fig. 2a and 2b represent the number of the day after the initial date although we use intraday data. We do that in order to simplify the discussion below. We select for both indices several time intervals with the following characteristics:

a) Intervals just before the crashes: the initial point is selected after the onset of the bubble and the last point is that of the all-time high of the index.
b) Intervals where no financial turbulences are observed.

For the NASDAQ we select three periods: from October 13 1995 to May 14 1997, from December 14 1998 to October 28 1999 and from October 28 1999 to February 24 2000. We labeled these periods as N1, N2, N3 respectively. In the period N1 no financial turbulence was observed, N2 corresponds to some important season of the Microsoft trial and N3 is just previous interval to the crash of April 2000 when NASDAQ loss about 25%.

For the IPC we also select three periods: from January 6 1994 to March 6 1995, from January 9 1996 to August 13 1997 and from August 13 1997 to October 25 1999. We labeled these periods as I1, I2, I3. The period I1 was disastrous for Mexican financial market due to the crisis brought about by the presidential transition[2]. In the period I2 no financial uncertainty was observed in economy and period I3 was highly turbulent due to the Asian crisis. We would like to stress the difference between I1 (endogenous origin of the crisis) and I3 (exogenous origin of the crisis).

We calculate $C(l)$ for the binary sequences associated to the above-mentioned intervals. The results appear in Fig. 3. Notice that the values of the function $C(l)$ corresponding to those intervals where no disturbance is observed (N1 and I2) are lower than those placed just before the crashes (N2, N3, I1, I3). It means that sequences corresponding to critical periods have more binary digits meaningful with respect to the whole series than those sequences corresponding to periods where nothing happened. The conclusion is that the intervals where no financial turbulence is observed, that is, where the markets works fine the informational contents of the binary series is small. If we compare the Fig. 3b with the Fig. 1 of [11] we conclude that N1 behave almost as a random sequence. This fact is close related with the EMH, because if all information about the market have been absorbed by the agents, then the behavior of prices will be random. It is a remarkable fact that those intervals where there is high uncertainty in the market, the information available is not fully incorporated by the agents and the informational content of the binary sequences is higher.

In the paper [12] we proposed an ansatz of the type $C(l) = dl^a$. The corresponding values of $d$ and $a$ for the sequences N1, …, I3 appear in the Table I. The larger exponents correspond to sequences with high financial turbulence. Notice that shorter exponents are related with the more random sequences. Therefore a good measure of how close is a market to the ideal situation described by the EMH is the exponent of the ansatz of $C(l)$.

---

[2] It was called "the tequilazo".

**4 Conclusions.**

The results of the last Section suggest that the intervals where the markets work fine produce binary sequences with features close to the random ones, meanwhile intervals with high uncertainty produce binary sequences, which carry more information.

The above is in agreement with the EMH, which stated that the markets are highly efficient in the determination of the most rational price of the traded assets. We conclude that a measure of how close is a markets to the ideal situation described by the EMH is the exponent $a$. The lower exponent, the more random sequence.

More surprisingly is the fact that intervals with high financial turbulence have high informational content. It open the challenges of understand what kind of information the sequence bears and how it would be used to predict the crashes.

**References.**

[1] R. N. Mantenga and H. E. Stanley, Econophysics, Cambridge University Press (1999).

[2] For a collection of papers and preprints on Minority Game see the web site:

http://www.unifr.ch/econophysics.

[3] D. Chalet, Y. C. Zhang, Phys. A **246**, 407 (1997).

[4] Y, C. Zhang, Europhys. News **29**, 51 (1998).

[5] R. Savit *et al.*, Phys. Rev. Let. **82**, 2203 (1999).

[6] R. Savit *et al.*, http://xxx.lanl.gov/abs/adap-org/9712006.

[7] N. F. Johnson *et al.*, Phys. A **256**, 230 (1998).

[8] A. Cavagna, http://xxx.lanl.gov/abs/cond-mat/9812215.

[9] N. F. Johnson *et al.*, http://xxx.lanl.gov/abs/cond-mat/9903164.

[10] D., Challet, Y. C. Zhang, Phys. A **256**, 514 (1998).

[11] R. Mansilla**,** Physical Review E, **62**, 4, 4553 (2000).

[12] R. Mansilla, Physica A **248**, 478 (2000).

[13] R. Mansilla, Comp. Syst. **11**, 387 (1997).

[14] W. H. Zurek, Nature **341**, 119 (1984).

[15] W. H. Zurek, Phys. Rev. A **40**, 4731 (1989).

[16] C. Adami, N. J. Cerf, http://xxx.lanl.gov/abs/adap-org/9605002.

[17] A. N. Kolmogorov, Rus. Math. Sur. **38**, 29 (1983).

[18] G. J. Chaitin, J. ACM **13**, 547 (1966).

[19] P. A. Samuelson, Ind. Man. Rev. **6**, 41 (1965).

[20] J. D. Farmer, http://xxx.lanl.gov/abs/adap-org/9812005

[21] A. Johansen, O. Ledoit and D. Sornette, Int. J. Theor. App. Fin. **3**, 2 (2000)

[22] B. G. Malkiel, A Random Walk Down Wall Street (W.W. Norton & Company, New York) (2000).

[23] G. K. Zipf, in Human Behavior and the Principle of Least Effort (Addison- Wesley, Cambridge, MA) (1949).

[24] N. Vandewalle, Ph. Boverox, F. Brisbois, http://xxx.lanl.gov/abs/cond-mat/0001293.

[25] J.C.G. Everetsz, Proc. 1$^{st}$ Int. Conf. on High-Freq. Data in Finance, Zurich, (1995).

**Figure captions.**

**Fig. 1**: A schematic representation of some scenarios in Minority game and in real financial markets. In Minority Game we always have situations as those labeled by C and D. In real financial markets we could have situations as A and B.

**Fig. 2**: The evolution of the NASDAQ (Fig. 2a) index and IPC (Fig. 2b) index in the time period from January 1995 to April 2000 and January 1991 to March 2000 respectively. The numbers which appear in the horizontal axis represent the number of the day after the initial date although we use intraday data. We do that in order to ease the comprehension. The intervals under study are: N1=(200, 360), N2=(1000, 1220) and N3=(1220, 1300), I1=(500, 790), I2=(180, 500) and I3=(1400, 1750).

**Fig. 3**: Values of $C(l)$ vs. $l$ for the intervals of interest (see text). In Fig. 3a we have I1(*) I2($\nabla$) and I3=(o). In Fig. 3b we have N1=($\nabla$), N2=(o), N3=(*). Notice that the $C(l)$ function for N1 behave as that of the random sequence (see Fig. 1 of reference [11]).

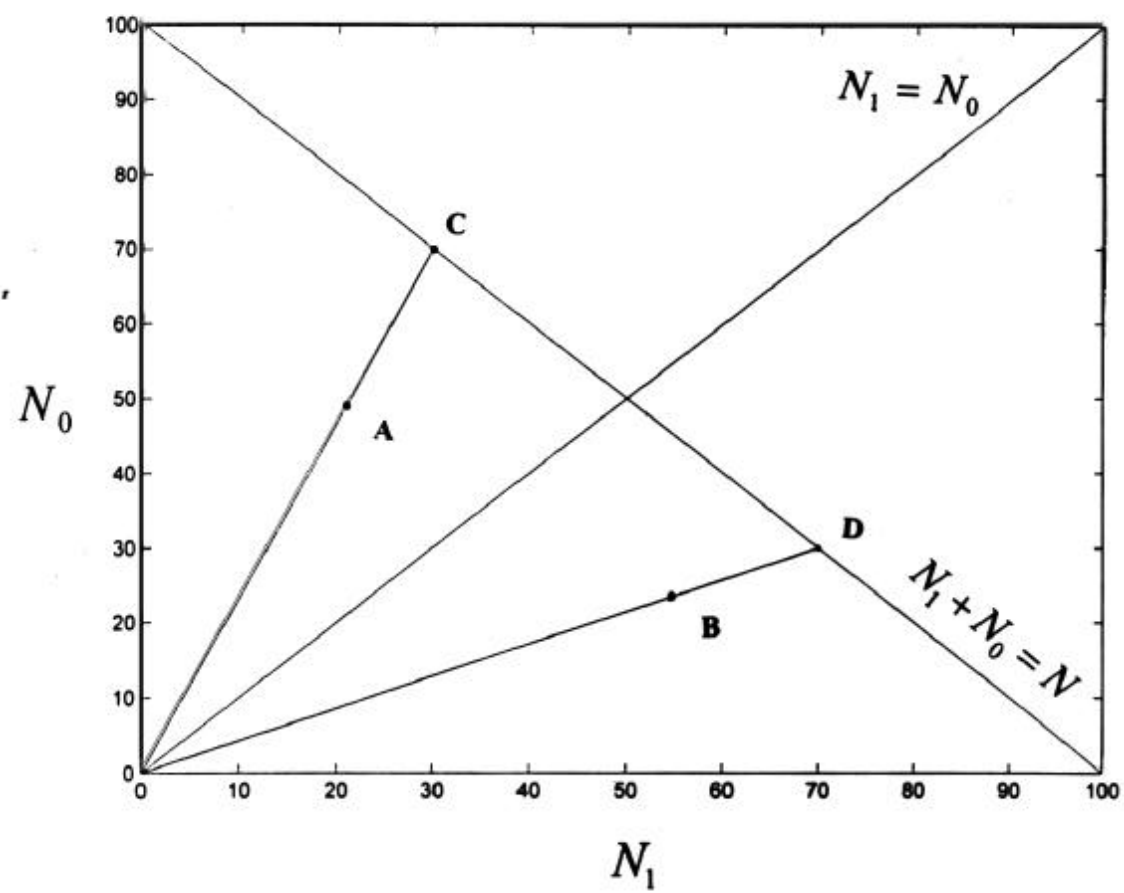

**Fig. 1**

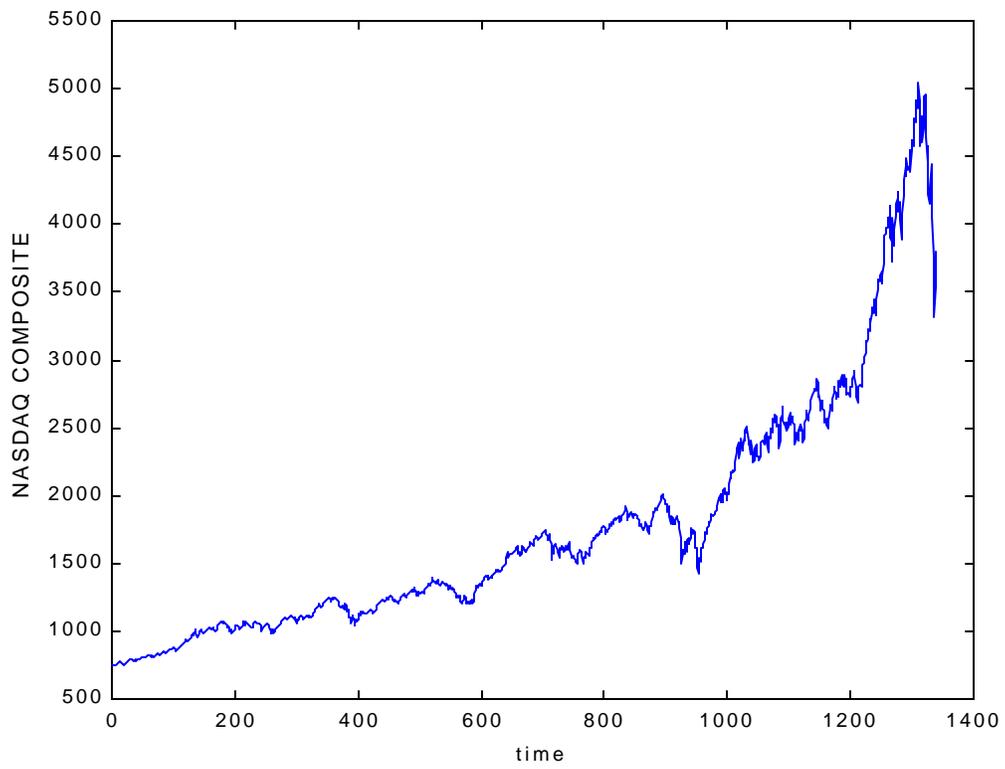

**Fig. 2a**

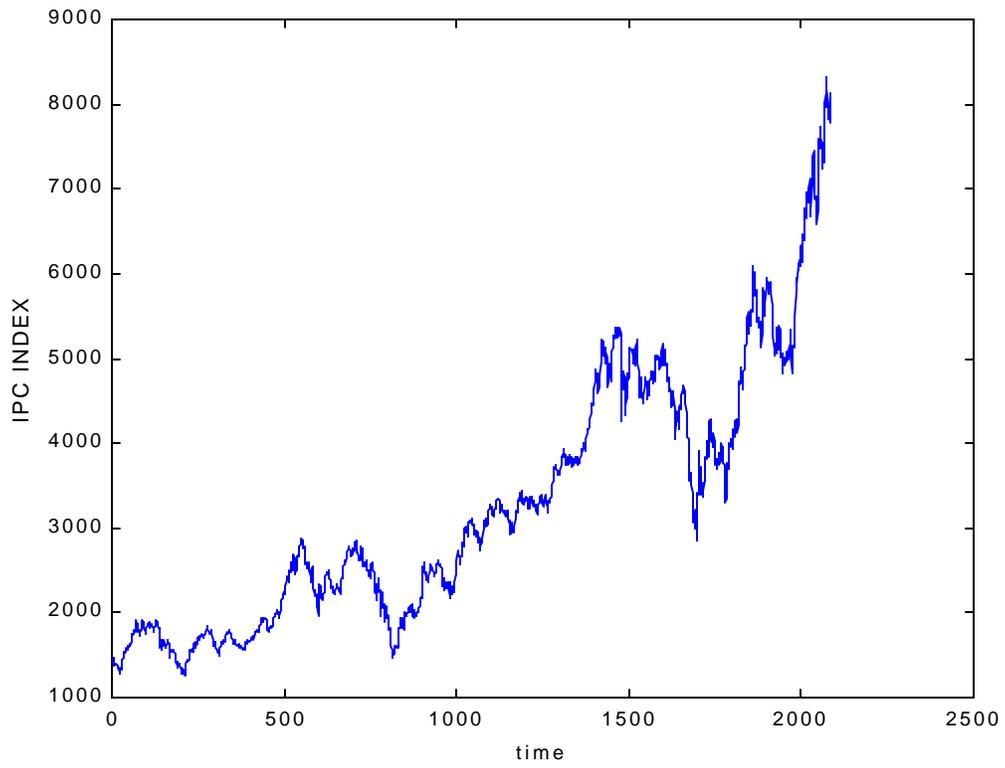

**Fig. 2b**

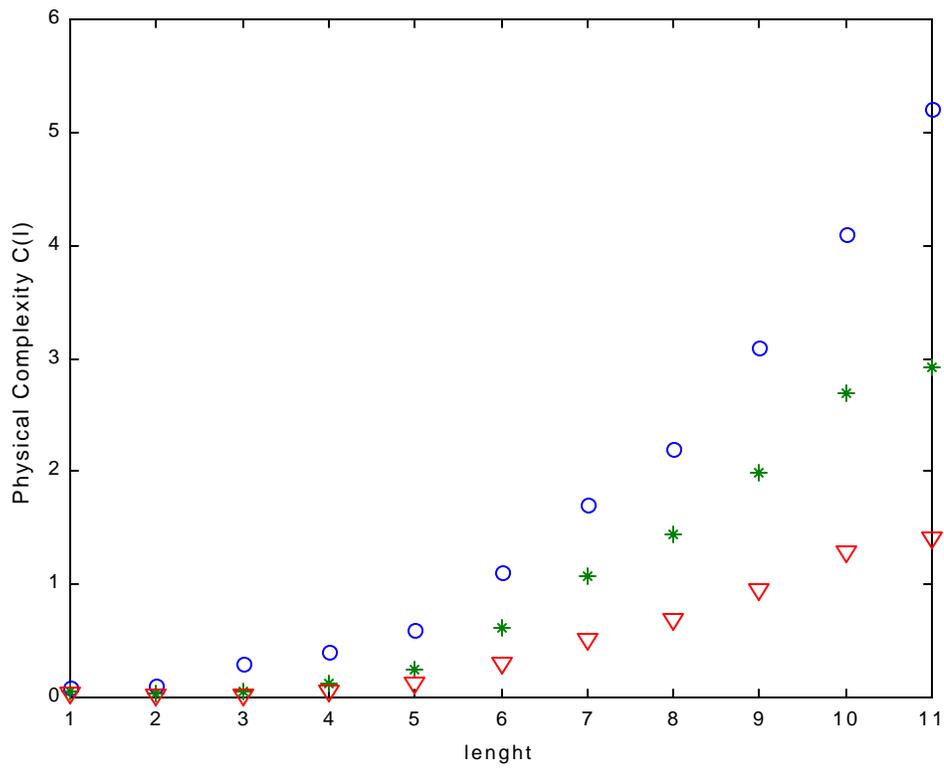

**Fig. 3a**

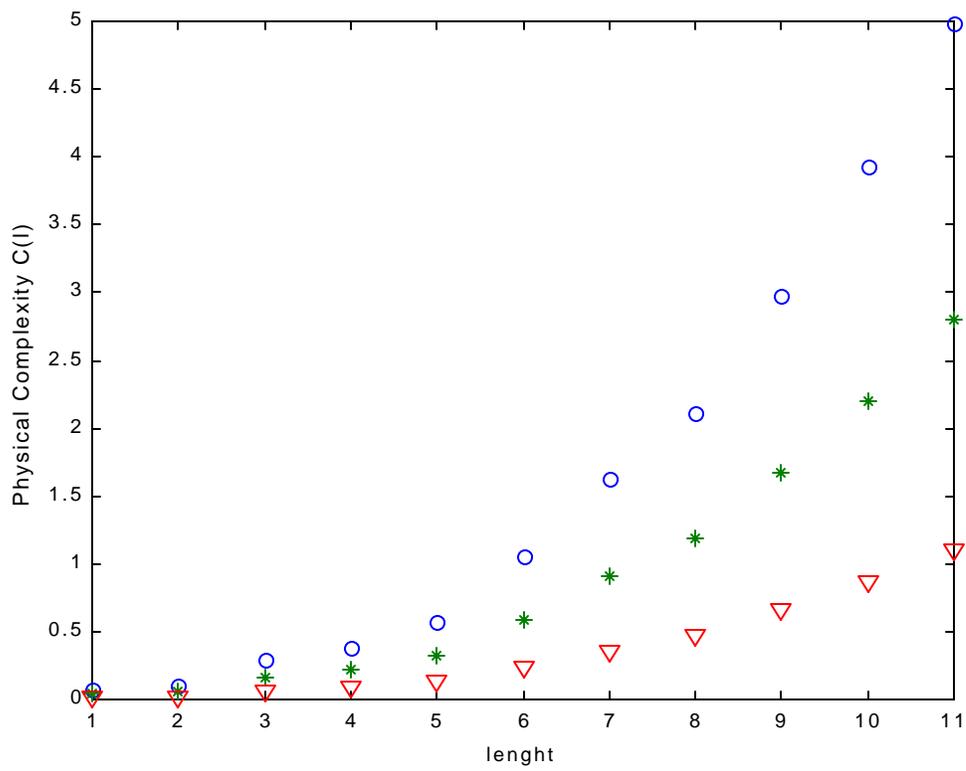

**Fig. 3b**

## Table I

## Values of *d* and *a* for the selected sequences.

| Sequences | a | d |
|-----------|--------|--------|
| N1 | 2.1059 | 0.0393 |
| N2 | 3.2458 | 0.0014 |
| N3 | 3.6715 | 0.0063 |
| I1 | 3.7748 | 0.0003 |
| I2 | 3.2213 | 0.0013 |
| I3 | 3.6901 | 0.0004 |